\documentclass[12pt]{article}
\pdfoutput=1

\usepackage{amsmath,amssymb,amsfonts,epsfig,cite,setspace,bigstrut,framed,eufrak}
\usepackage[all]{xy}
\usepackage{color}
\usepackage{pifont}

\makeatletter \@addtoreset{equation}{section} \makeatother
\renewcommand{\theequation}{\thesection.\arabic{equation}}

\begin{document}

\vskip 0.25in

\newcommand{\todo}[1]{{\bf\color{blue} !! #1 !!}\marginpar{\color{blue}$\Longleftarrow$}}
\newcommand{\nn}{\nonumber}
\newcommand{\comment}[1]{}
\newcommand\T{\rule{0pt}{2.6ex}}
\newcommand\B{\rule[-1.2ex]{0pt}{0pt}}

\newcommand{\CO}{{\cal O}}
\newcommand{\cI}{{\cal I}}
\newcommand{\cM}{{\cal M}}
\newcommand{\cW}{{\cal W}}
\newcommand{\cN}{{\cal N}}
\newcommand{\cR}{{\cal R}}
\newcommand{\cH}{{\cal H}}
\newcommand{\cK}{{\cal K}}
\newcommand{\cT}{{\cal T}}
\newcommand{\cZ}{{\cal Z}}
\newcommand{\cO}{{\cal O}}
\newcommand{\cQ}{{\cal Q}}
\newcommand{\cB}{{\cal B}}
\newcommand{\cC}{{\cal C}}
\newcommand{\cD}{{\cal D}}
\newcommand{\cE}{{\cal E}}
\newcommand{\cF}{{\cal F}}
\newcommand{\cA}{{\cal A}}
\newcommand{\cX}{{\cal X}}
\newcommand{\IA}{\mathbb{A}}
\newcommand{\IP}{\mathbb{P}}
\newcommand{\IQ}{\mathbb{Q}}
\newcommand{\IH}{\mathbb{H}}
\newcommand{\IR}{\mathbb{R}}
\newcommand{\IC}{\mathbb{C}}
\newcommand{\IF}{\mathbb{F}}
\newcommand{\IV}{\mathbb{V}}
\newcommand{\II}{\mathbb{I}}
\newcommand{\IZ}{\mathbb{Z}}
\newcommand{\re}{{\rm Re}}
\newcommand{\im}{{\rm Im}}
\newcommand{\tr}{\mathop{\rm Tr}}
\newcommand{\ch}{{\rm ch}}
\newcommand{\rk}{{\rm rk}}
\newcommand{\ext}{{\rm Ext}}
\newcommand{\bi}{\begin{itemize}}
\newcommand{\ei}{\end{itemize}}
\newcommand{\beq}{\begin{equation}}
\newcommand{\eeq}{\end{equation}}
\newcommand{\bea}{\begin{eqnarray}}
\newcommand{\eea}{\end{eqnarray}}
\newcommand{\ba}{\begin{array}}
\newcommand{\ea}{\end{array}}

\newcommand{\CN}{{\cal N}}
\newcommand{\y}{{\mathbf y}}
\newcommand{\z}{{\mathbf z}}
\newcommand{\C}{\mathbb C}\newcommand{\R}{\mathbb R}
\newcommand{\CA}{\mathbb A}
\newcommand{\CP}{\mathbb P}
\newcommand{\cP}{\mathcal P}
\newcommand{\tmat}[1]{{\tiny \left(\begin{matrix} #1 \end{matrix}\right)}}
\newcommand{\mat}[1]{\left(\begin{matrix} #1 \end{matrix}\right)}
\newcommand{\diff}[2]{\frac{\partial #1}{\partial #2}}
\newcommand{\gen}[1]{\langle #1 \rangle}

\newtheorem{theorem}{\bf THEOREM}
\newtheorem{proposition}{\bf PROPOSITION}
\newtheorem{observation}{\bf OBSERVATION}

\def\theequation{\thesection.\arabic{equation}}
\newcommand{\setall}{
	\setcounter{equation}{0}
}
\renewcommand{\thefootnote}{\fnsymbol{footnote}}

\begin{titlepage}
\vfill
\begin{flushright}
{\tt\normalsize KIAS-P17009}\\

\end{flushright}
\vfill

\begin{center}
{\Large\bf D-Particles on Orientifolds and Rational Invariants}

\vskip 1.5cm

Seung-Joo Lee\footnote{\tt seungsm@vt.edu} and
Piljin Yi\footnote{\tt piljin@kias.re.kr}
\vskip 5mm
{\it $^*$Department of Physics, Robeson Hall, Virginia Tech, \\
Blacksburg, VA 24061, U.S.A.
}
\vskip 3mm
{\it $^\dagger$School of Physics,
Korea Institute for Advanced Study, Seoul 02455, Korea}

\end{center}
\vfill

\begin{abstract}
We revisit the D0 bound state problems,  of
the M/IIA duality, with the Orientifolds.
The cases of O4 and O8 have been studied recently, from the
perspective of five-dimensional theories, while the case
of O0 has been much neglected. The computation we perform for
D0-O0 states
boils down to the Witten indices for ${\cal N}=16$ $O(m)$
and $Sp(n)$ quantum mechanics, where we  adapt and
extend previous analysis by the authors. The
twisted partition function $\Omega$, obtained via localization,
proves to be {\it rational}, and we establish a precise relation
between $\Omega$ and the {\it integral} Witten index $\cI$,
by identifying continuum contributions sector by sector.
The  resulting Witten index shows surprisingly large numbers
of threshold bound states but in a manner consistent with M-theory.
We close with an exploration on how the ubiquitous rational
invariants of the wall-crossing physics would generalize
to theories with Orientifolds.

\end{abstract}

\vfill
\end{titlepage}

\tableofcontents
\renewcommand{\thefootnote}{\#\arabic{footnote}}
\setcounter{footnote}{0}
\vskip 2cm

\section{Introduction}

One of the earliest BPS bound state counting problems in the context of
superstring theory is that of multi-D0 threshold bound states.
M theory/IIA theory duality anticipates supersymmetric bound
states of $N$ D-particles, for all natural numbers $N$ \cite{Witten:1995ex}.
This problem was given a lot of attention since its first inception
by Witten, and obviously, $N=2$ case, i.e., $\cN=16$ $SU(2)$
quantum mechanics, has been dealt with the most rigor
\cite{Yi:1997eg,Sethi:1997pa}, while higher $N$ cases
have lead to many new insights over the years.

This problem was given a fresh treatment recently via
the localization technique \cite{HKY,Lee:2016dbm}. Previously,
computation of twisted partition functions had been performed for
$\cN=16$ $SU(N)$ theories \cite{Moore:1998et} and some attempts made
for other gauge groups \cite{Staudacher:2000gx,Pestun:2002rr},
but there are often issues with a contour choice in the last stage of such computations.
The new derivations obviate this last uncertainty as they
actually derive rigorously what the contour should be.
For $SU(N)$, one finds in the end the twisted partition function \cite{Lee:2016dbm}
\bea
\Omega^{SU(N)}_{\cN=16}= 1+\sum_{p\vert N; p>1}1\cdot \Delta^{SU(p)}_{\cN=16}\ ,
\eea
with rational functions $\Delta$ whose precise form
for a general Lie Algebra can be found in Eq.~\eqref{Delta16}.

This $\Omega$, being non-integral, is certainly not the same as
the Witten index \cite{Witten:1982df}. Such
is usually a symptom of having asymptotic flat directions
that cannot be lifted by a parameter tuning. For $SU(N)$
theory in question, the classical vacua form a cone
${\mathbb R}^{9(N-1)}/S_N$, and the plane-wave-like
states can also contribute  to the relevant path integral. The
correct interpretation here is to identify the first term ``1"
as the index while the rest are attributed to various continuum
sectors. In fact, the other ``1"'s in the sum are also
nothing but the Witten index of the $SU(N/p)$ subsectors.
This interpretation was pioneered in Ref.~\cite{Yi:1997eg},
where the nonequivariant version of $\Omega$ was computed
for $SU(2)$, and has been generalized to all $SU(N)$ rather
convincingly \cite{Green:1997tn,Moore:1998et}.

Thus one question that has to be resolved if one is to repeat
the problem for more complicated spacetime is how to separate
the continuum contribution from true Witten index systematically.
This does not seem to admit a universal answer, as
there are numerous cases where continuum sectors can conspire
to contribute a net integral piece to $\Omega$ \cite{Lee:2016dbm}.  At present,
extraction of $\cI$ from $\Omega$, when the theory involves
gapless asymptotic directions, is more of an art than a science.

Ref.~\cite{Lee:2016dbm}, nevertheless, noted how the main
feature of $\cN=16$ $SU(N)$ generalizes straightforwardly
to other $\cN=16$ theories and also to $\cN=4$ non-primitive
quiver theories with bifundamental matters only.
The various continuum contributions to $\Omega^G_{\cN=16}$ have been
physically understood, identified, and catalogued. Naturally,
this  opens up the possibility of computing true Witten
indices for D-particle binding to Orientifold points.
In fact, the results of Ref.~\cite{Lee:2016dbm} almost suffice,
except for the case of O0$^-$ orientifold. In this note, we wish
to place the last missing piece in the problem and
compute Witten indices for all D0-O0 bound states.

Section 2 will give a general discussion on the twisted
partition function versus the Witten index, with emphasis
on what the localization procedure actually computes.
Section 3 will review the recent results for $\cN=16$
Yang-Mills quantum mechanics, which we will generalize
in Section 4 to $O(m)$ gauge groups. This will lead us
to the Witten indices that count bound states between D0's and
any one of four types of the orientifold point and
to a known M-theory interpretation, adding yet another strong
and rather direct confirmation of M/IIA duality. In the
final section, we comment on new type of rational
expressions we found along the way and propose them
as building blocks for the rational invariants suitable
for Orientifolded theories.

\section{Index ${\cal I}$ vs. Twisted Partition Function $\Omega$ }

For supersymmetric quantum theory, one of the useful and accessible
quantities that probe the ground state sector is the Witten index \cite{Witten:1982df},
$$
\cI=\lim_{\beta\rightarrow \infty}{\rm tr}
\left[ (-1)^{\cal F} e^{-\beta H}\right] \ .
$$
The chirality operator  $(-1)^{\cal F}$ can be replaced by any operator that anti-commutes
with the supercharges. One often wishes to compute the equivariant
version by inserting chemical potentials, $x$, associated with global
symmetries, $F$,
$$
\cI(x)\equiv \lim_{\beta\rightarrow \infty}{\rm tr}
\left[ (-1)^{\cal F} x^{F} e^{-\beta H}\right] \ .
$$
Even more useful information emerges if we select out a particular
supercharge ${\cal Q}$ which commutes with a linear combination of
$R$-symmetry generators, call it $R$, and one of the $F$'s,
resulting in a fully equivariant Witten index,
\bea\label{WI}
\cI({\bf y},x)\equiv \lim_{\beta\rightarrow \infty}{\rm tr}
\left[ (-1)^{\cal F} {\bf y}^{R} x^{F} e^{-\beta {\cal Q}^2}\right] \ .
\eea
However, as is well known, this quantity may not be amenable
to straightforward computations.

If the dynamics is compact, i.e., with a fully
discrete spectrum, $\beta$-dependence can be argued away
based on the naive argument that ${\cI}$ is topological.
Under such favorable circumstances, one is motivated to consider
instead
\bea
\Omega({\bf y},x,\beta)\equiv {\rm tr}
\left[ (-1)^{\cal F} {\bf y}^{R} x^{F} e^{-\beta {\cal Q}^2}\right] \ ,
\eea
and compute the other limit, which tends to reduce the
path integral to a local expression,
\bea
{\cal I}_{\rm bulk}({\bf y},x)\equiv\lim_{\beta\rightarrow0}\Omega({\bf y},x,\beta) \ ,
\eea
with the anticipation that $\Omega$ is independent of $\beta$ so
that ${\cal I}={\cal I}_{\rm bulk} $.

For theories with continuum sectors, however, this naive
expectation cannot hold in general; ${\cal I}$ is by
definition integral, while $\Omega$ need not be integral
and thus can differ from ${\cal I}$. If the continuum
has a gap, $E\ge E_{\rm gap}>0$, its contribution
is suppressed as
$$ e^{-\beta E_{\rm gap}} \ , $$
so we may have an option of scaling $E_{\rm gap}$ up first
and then taking $\beta\rightarrow 0$ afterward, leaving
behind the integral index $\cI$ only \cite{HKY,Stern:2000ie}.

When the continuum cannot be gapped, or when a gap can be
introduced only at the expense of qualitative modification of
the asymptotic dynamics, however, we are often in
trouble. The resulting bulk part ${\cal I}_{\rm bulk}$ differs
from the genuine index. For such theories, isolating $\cI$
hidden inside $\cI_{\rm bulk}$ requires a method
of computing yet another piece, known as the defect term,
\bea
-\delta\cI
\equiv
\cI_{\rm bulk}-\cI \ .
\eea
This program depends on particulars of the
given problem and, in particular, on the
boundary conditions.\footnote{
A canonical example is the supersymmetric
nonlinear sigma models onto a manifold with boundary. If one
adopt the so-called APS boundary condition, $\delta \cI$ is
then computed by the eta-invariant, leading to the Atiyha-Patodi-Singer
index theorem \cite{APS}. This boundary condition, however, does not
in general translate to $L^2$ condition on the physical space.}
As far as we know there is no general theory for $\delta \cI$.

For a large class of gauged dynamics, the localization
procedure has been applied successfully to reduce the path
integral representation of $\Omega$
to a formulae involving rank-many contour integrations.
For ${\cal N}\ge 2$
gauged quantum mechanics \cite{HKY} and for $d=2$ elliptic
genera
\cite{Benini:2013nda,Benini:2013xpa}, in particular,
reasonably complete and reliable derivations exist.
At the end of such computations, one finds that
$\beta$-dependence is absent. When the dynamics is
not compact and $\Omega$ is expected to be
$\beta$-dependent, the question is exactly which
$\beta$ limit of $\Omega$ one has computed.

One key trick here is to scale up the gauge kinetic term
by sending $e^2\rightarrow 0$, as the term is often BRST-exact
for the spacetime dimension $D$ less than three. In the absence
of other dimensionful parameters of the theory,
the only obvious answer to the question we posed above
is $\beta\rightarrow0$; The dimensionless combination
of the two is
$$
\beta e^{2/(4-D)} \ ,
$$
so $e^2\rightarrow 0$ is equivalent to $\beta\rightarrow 0$
for $D\le 3$. Another typical dimensionful parameters that
could be present are Fayet-Iliopoulos constants $\zeta$,
but, for a sensible results, one often must take a
limit of $\zeta$ first \cite{HKY}. This raises a gap
$E_{\rm gap}$ along certain Coulomb directions to infinity,
if not all, so we expect that, again, the $\beta\rightarrow 0$ limit
of $\Omega$ is computed effectively
at the end of the localization procedure. After all,
one finds a local expression, at the end of such processes,
involving zero mode integrals only, which is impossible at the other
limit of $\beta\rightarrow \infty$.

As such, we will define for this note,
\bea
\Omega({\bf y},x) \equiv \Omega({\bf y},x,\beta)\biggr\vert_{\rm localization}\ ,
\eea
whereby, according to the above scaling argument, we may identify
\bea
\cI_{\rm bulk}=\Omega({\bf y},x) \ .
\eea
We will call this quantity the twisted partition function,
although, strictly speaking, the true twisted partition
function $\Omega({\bf y},x,\beta)$ may yet differ from
$\Omega({\bf y},x)$. This brings us to a general statement
\bea
\cI({\bf y},x)=\Omega({\bf y},x) +\delta \cI({\bf y},x) \ .
\eea
Even after a successful localization computation
of $\Omega({\bf y},x)$, one is often left with an even more difficult
task of identifying the continuum contribution, $-\delta \cI$,
inside $\Omega$ if one wishes to compute $\cI$.

There appears to be no single universal relationship
between $\cI$ and $\Omega$, but surprisingly, as delineated
in Ref.~\cite{Lee:2016dbm}, there exists classes of $d=1$
supersymmetric gauged linear sigma models for which this problem
may be dealt with honestly. One such  is adjoint-only
Yang-Mills quantum mechanics, and another is $\cN=4$  nonprimitive
quiver theories with compact classical Higgs vacuum moduli space.
In the next section, we recall this phenomenon for
$\cN = 4,8,16$ pure Yang-Mills quantum mechanics
with connected simple group $G$.

\section{Rational $\Omega^G_\cN $ and Integral $\cI^G_\cN$}

For gauged linear sigma model with at least two supersymmetries,
the localization procedure gives a Jeffrey-Kirwan residue
formulae \cite{HKY},
\bea\label{jk-formula}
\Omega({\bf y},x)
=\frac{1}{|W|}{\text{JK-Res}}_\eta\;\frac{g(t)}{\prod_s t_s}\;{\rm d}^r t \ ,
\eea
where $(t_1,\dots,t_r)$ parameterize the $r$ bosonic zero modes
living in $({\mathbb C}^*)^{r}$, that usually scan the Cartan
directions but can be further restricted in topologically  nontrivial
holonomy sectors. The determinant $g(t)$ is due to massive
modes in the background of $t$'s.
In this note, we use $\cN=4$ notations for supermultiplets,
and as such, $g(t)$ takes the general form,
\bea
g(t)
&=&\left(\frac{1}{{\bf y}-{\bf y}^{-1}}\right)^{r}
\prod_{\alpha}\frac{t^{-{\alpha/ 2}}-t^{\alpha/ 2}}{
t^{\alpha/ 2}{\bf y}^{-1}-t^{-{\alpha/ 2}}{\bf y}} \cr\cr\cr
&&\times \prod_{i}\frac{t^{-Q_i/2}x^{-{F_i/ 2}}{\bf y}^{-\left({R_i/ 2}-1\right)}
-t^{Q_i/2}x^{{F_i/ 2}}{\bf y}^{{R_i/ 2}-1}}{
t^{Q_i/2}x^{{F_i/ 2}}{\bf y}^{R_i/ 2}
-t^{-Q_i/2}x^{-{F_i/ 2}}{\bf y}^{-{R_i/ 2}}} \ .
\label{4g}
\eea
Here, $\alpha$ runs over the roots of the gauge group and $i$
labels the individual chiral multiplets, with the gauge
charge $Q_i$ and the flavor charge $F_i$ under the Cartans
of the gauge group and of the flavor group, respectively.
Finally, $W$ is the Weyl group of the gauge group and $\eta$
is a choice of $r$ auxiliary parameters. For detailed definition of the
JK residue \cite{JK}, the condition on the auxiliary parameters $\eta$,
and the derivation of the above formula, we refer the
reader to the section 4 of Ref.~\cite{HKY}. We will refer to
this general procedure as HKY.

For pure $\cN=4$ theories, the computation admits the R-charge
chemical potential ${\bf y}$ only. For ${\cal N}=8,16$, we have
additional adjoint chirals, and the assignment of global
charges needs a little bit of thought. For $\cN=8$, one more
chemical potential $x$ can be turned on, associated with the
natural $U(1)$ rotation of the chiral field, and $R=0$ is
assigned to the adjoint chiral. No superpotential is possible
under such assignments.
For $\cN=16$, with three adjoint chirals, a trilinear
superpotential term is needed, so at most two flavor
chemical potentials are allowed, say, $x$ and $\tilde x$
associated with $F$ and $\tilde F$. We can
for example assign $R=(2,0,0)$, $F=(2,-1,-1)$, and
$\tilde F=(0,1,-1)$ that allow only trilinear superpotential
as required by $\cN=16$. In actual $\cN=16$ formula below $x^{F}$ should
be understood as the product, $x^{F}\tilde x^{\tilde F}$, over
the two flavor chemical potentials.

One thing special about the pure gauge theories is that
we are instructed to ignore the poles located at the
boundary of the zero mode space $({\mathbb C}^*)^r$ \cite{Lee:2016dbm}.
This is a property which holds generally for theories
with the total matter content in a real representation under
the gauge group.

\subsection{${\cal N}=4,8$ }

This gives us an unambiguous procedure of computing the twisted
partition functions $\Omega^G_\cN$ for all possible $G$ and $\CN$.
There are some further computational issues, such as how to
deal with the degenerate poles, which complicates the task
but still allows us to go forward. We will not give too much
details here and instead refer the readers to Ref.~\cite{Lee:2016dbm}
for pure Yang-Mills cases, and to Ref.~\cite{HKY} for general
gauged quantum mechanics.

It turns out that, after a long and arduous computer-assisted
computation of JK residues, the twisted partition functions for
pure $\CN=4,8$ $G$-gauged quantum mechanics, can be
organized into purely algebraic quantities. For $\cN=4$, one finds
\bea\label{N=4}\label{N=4}
\Omega^G_{\CN=4}({\bf y})=
\frac{1}{|W_G|}\sum'_{w}\frac{1}{{\rm det}\left({\bf y}^{-1}-{\bf y}\cdot w\right)}\ .
\eea
The sum is only over the elliptic Weyl elements and
$|W_G|$ is the cardinality of the Weyl group itself. An elliptic
Weyl element $w$ is defined by absence of eigenvalue 1;
In other words, in the canonical $r$-dimensional representation
of the Weyl group on the weight lattice,
$${\rm det}\left(1-w\right)\neq 0\ .$$
Some simple examples with $\cN=4$ are
\bea
\Omega_{\cN=4}^{SU(N)}({\bf y})&=&\frac{1}{N}\cdot \frac{1}{{\bf y}^{-N+1}+{\bf y}^{-N+3}+\cdots+{\bf y}^{N-3}+{\bf y}^{N-1}}\ , \cr\cr
\Omega_{\cN=4}^{SO(4)}({\bf y})&=&\frac14\cdot \frac{1}{({\bf y}^{-1}+{\bf y})^2}\ , \cr\cr
\Omega_{\cN=4}^{SO(5)}({\bf y})=\Omega_{\cN=4}^{Sp(2)}({\bf y})&=&
\frac{1}{8}\cdot\left[\frac{2}{{\bf y}^{-2}+{\bf y}^2}+\frac{1}{({\bf y}^{-1}+{\bf y})^2}\right]\ ,\cr\cr
\Omega_{\cN=4}^{SO(7)}({\bf y})=\Omega_{\cN=4}^{Sp(3)}({\bf y})&=&
\frac{1}{48}\cdot\left[\frac{8}{{\bf y}^{-3}+{\bf y}^3}+ \frac{ 6 }{({\bf y}^{-2}+{\bf y}^2)({\bf y}^{-1}+{\bf y})}
+\frac{1}{({\bf y}^{-1}+{\bf y})^3}\right]\ ,\nonumber
\eea
where each term can be associated with a sum over
conjugacy classes of the same cyclic decompositions.

For pure ${\cal N}=8$ $G$-gauged quantum mechanics, obtained by adding
to the $\CN=4$ theory an adjoint chiral, we can include a flavor
chemical potential $x$ of the adjoint after assigning a unit
flavor charge without loss of generality. With $R=0$ for
the adjoint chiral, we also have the universal formula,
\bea\label{N=8}
\Omega^G_{\CN=8}({\bf y},x)=\frac{1}{|W_G|}\sum'_{w}\frac{1}{{\rm det}
\left({\bf y}^{-1}-{\bf y}\cdot w\right)}\cdot
\frac{{\rm det}\left({\bf y}^{-1} x^{1/2}-{\bf y} x^{-1/2}\cdot
w\right)}{{\rm det}\left( x^{1/2}- x^{-1/2}\cdot w\right)}\ ,
\eea
where again the sum is over the elliptic Weyl elements of $G$. For
example we have
$$\Omega_{\cN=8}^{SO(4)}({\bf y},x)=
\frac14\cdot \frac{1}{({\bf y}^{-1}+{\bf y})^2}\cdot
\frac{({\bf y}^{-1}x^{1/2}+{\bf y}x^{-1/2})^2 }{(x^{1/2}+x^{-1/2})^2}\ ,
$$
and the pattern generalizes to higher rank cases in an obvious manner.

The reason why the result can be repackaged into such a
simple algebraic formulae has been explained
both for nonequivariant form \cite{Yi:1997eg,Green:1997tn,Kac:1999av,Kim:2011sc}
and for equivariant form \cite{Lee:2016dbm}. Consider $-\delta \cI$.
This part of $\Omega $ has to arise from
the continuum and, because of this, depends only on the asymptotic dynamics.
The latter becomes a nonlinear sigma model on an orbifold
\bea
\CO(G)_{\cN=4,8}=\IR^{3r}/W\;{\rm or}\;\IR^{5r}/W \ ,
\eea
so that the $\delta\cI$ of the two theories must agree with
each other. On the other hand, we expect no quantum mechanical
bound state localized at the orbifold point, so
$$ \left(\cI^{\CO(G)}_{\cN=4,8}\right)_{\rm bulk}  +
\delta\cI^{\CO(G)}_{\cN=4,8}= 0$$
which implies~\cite{Yi:1997eg}
\bea\label{defect}
-\delta{\cI}^G_{\cN=4,8} =\left(\cI^{\CO(G)}_{\cN=4,8}\right)_{\rm bulk}\ .
\eea
The right hand side of (\ref{defect})
has been evaluated using the Heat Kernel regularization, when
${\bf y}=1$ and $x=1$, for $SU(2)$ case in Ref.~\cite{Yi:1997eg},
and more generally in Refs.~\cite{Green:1997tn,Kac:1999av}, with
the result
\bea
\frac{1}{|W|}\sum'_{w}\frac{1}{{\rm det}\left(1- w\right)} \ .
\eea
What we described above in (\ref{N=4}) and in (\ref{N=8}), individually confirmed
by direct localization computation, are the
equivariant uplifts of this expression for $\cN=4,8$ respectively.

With this, the origin of $\Omega^G_{\cN=4,8}$ is abundantly
clear. They come entirely from the asymptotic continuum states
spanned by the free Cartan dynamics, modulo the orbifolding
by the Weyl group; The path-integral-computed $\Omega^G_{\cN=4,8}$
has no room for a contribution from threshold bound states. Therefore, the true enumerative
part $\cI$ inside $\Omega$ has to be null,
\bea
\cI^G_{\cN=4}=0= \cI^G_{\cN=8}\ ,
\eea
for any simple group $G$. Recall that, for classical groups $G$, $\cN=4,8$ pure
Yang-Mills quantum mechanics has no bound state, as can be
argued based on D2/D3-branes multiply-wrapped on $S^2$ and $S^3$
in K3 and Calabi-Yau  three-fold, possibly together with Orientifold
planes, and the Witten index of these theories must vanish. This physical
expectation dovetails with the above structure nicely.

The same principle generalizes to $\cN=16$ cases. However,
their asymptotic dynamics will no longer be captured by analog of
$\cO(G)$ alone; The presence of threshold bound states implies that the
continuum sectors $\Omega_{\cN=16}^G$ will no longer be that
simple. There could be additional sectors involving partial
bound states tensored with continuum of remaining asymptotic
directions. We turn to this next.

\subsection{On ${\cal N}=16$ Continuum Sectors}

The same kind of continuum sectors as the above $\cN=4,8$
examples should exist for $\cN=16$, with the asymptotic
dynamics of the form,
\bea
\cO(G)_{\cN=16}={\mathbb R}^{9r}/W_G\ ,
\eea
and we can easily guess the contribution to
$\Omega^G_{\cN=16}$ from this sector to take the
form,
\bea\label{Delta16}
&&\Delta^{G}_{\cN=16}\equiv\\ \cr
&&\frac{1}{|W_{G}|}\sum'_{w}\frac{1}{{\rm det}
\left({\bf y}^{-1}-{\bf y}\cdot w\right)}\cdot
\prod_{a=1}^3
\frac{{\rm det}\left(x^{F_a/2}{\bf y}^{R_a/2-1}-x^{-F_a/2}{\bf y}^{1-R_a/2}
\cdot w\right)}{{\rm det}\left( x^{F_a/2}{\bf y}^{R_a/2}
- x^{-F_a/2}{\bf y}^{-R_a/2}\cdot w\right)}\ ,\nonumber
\eea
as a straightforward generalization of $\cN=4,8$ expressions.
Here, $a$ labels the three adjoint chirals. Indeed, as we will
see below, each $\Omega^G_{\cN=16}$, computed via localization,
is seen to have an additive piece of this type.

The difference for $\CN=16$ is, however, that threshold bound
states are expected in general. For all $SU(N)$, e.g.,
a single threshold bound state must exist for M-theory/IIA
theory duality to hold. Since such states can also occur
for subgroups of $G$ as well and since they
can explore the remaining asymptotic directions, a far more
complex network of continuum sectors exist. Generally a product
of subgroups
$$\otimes_A G_A \quad<\quad  G$$
correspond to a collection of one-particle-like states, each
labeled by $A$. When this subgroup equals the Cartan subgroup
of $G$, the corresponding continuum sector contributes
the universal $\Delta^G_{\cN=16}$ to $\Omega^G_{\cN=16}$.
When at least one of $G_A$ is a simple group, the corresponding
partial bound state(s) can contribute a new fractional piece to
$\Omega^G_{\cN=16}$. The relevant continuum sector is the
asymptotic Coulombic directions where the ``particles" forming
the bound state associated with $G_A$ moves together. In
other words, the asymptotic Coulombic directions are parameterized
by a subalgebra
$$h[\otimes_A G_A]$$
of the Cartan of $G$, where $\otimes_A G_A$ is the centralizer
of $h[\otimes_A G_A]$.

Then, the argument leading to (\ref{defect}) can be adapted
to this slightly more involved case; A continuum contribution
from this sector would be associated with a subgroup
$$W'$$ of $W_G$ that leaves $h[\otimes_A G_A]$ invariant yet act
faithfully. Contribution to $\Omega$ would arise from
generalized elliptic Weyl elements of $W'$,
$${\rm det}\left(1-w'\right)\biggr\vert_{h[\otimes_A G_A]}\neq 0\ ,$$
where the determinant is now taken in the smaller representation
over $h[\otimes_A G_A]$. In a slight abuse of notation,
it turns out that the continuum contribution from $W'$ to
$\Omega^G_{\cN=16}$ can be expressed as a product of the form,
$$\prod_I \Delta^{H_I}_{\cN=16}\, $$
where $\Delta^{H_I}_{\cN=16}$ are defined for some subgroups $H_I$
of $G$ in the same manner as (\ref{Delta16}). Each $H_I$ is a
simple subgroup of $G$ whose Weyl group is a subgroup factor of $W'$.

\subsection{${\cal N}=16$}

$\Omega^G_{\cN=16}$ can also be directly computed using the
HKY procedure \cite{HKY}. One then searches for a unique decomposition
as sum over such continuum pieces  as
\begin{equation}
\Omega^{G}_{\CN=16}
\;=\;\cI_{\CN=16}^{G}+
\sum_{\otimes {{G}_A}< {G}}n^G_{\{G_A\}} \prod_{I} \Delta^{{H}_I}_{\cN=16} \ ,
\end{equation}
with nonnegative integral factor, $n^G_{\{G_A\}}$. Furthermore,
there should be a term
$$1\cdot \Delta^{G}_{\cN=16}$$
on the right hand side, with the coefficient 1, representing
the sector with no partial bound state whatsoever.

Ref.~\cite{Lee:2016dbm} showed that this is indeed
the case, even though such a pattern is hardly visible
at the stage of JK-residue computations.
For $SU(N)$, the result takes a particularly simple form,
\begin{equation}\label{16}
\Omega^{SU(N)}_{\CN=16}\;=\;1+
\sum_{p\vert N; p\neq 1}1 \cdot  \Delta^{SU(p)}_{\cN=16}  \ .
\end{equation}
The rational contributions come from the continuum directions,
$h[\otimes_A G_A]$, parameterized as
$$
{\rm diag}(v_1,\dots, v_1; v_2,\dots ,v_2;\quad\cdots\quad ; v_p,\dots ,v_p)$$
with each eigenvalue repeated $(N/p)$-times, and $\sum_A v_A=0$.
In this sector, $p$ number of partial $SU(N/p)$ bound states
form, continuum states of which contribute $\Delta^{SU(p)}_{\cN=16}$;
The relevant Weyl subgroup is the permutation group that
shuffles $v$'s, so can be naturally labeled as $H=SU(p)$.
In the end, this implies
\bea\cI_{\cN=16}^{SU(N)}=1
\eea
for all $N$. The nonequivariant limit of the same decomposition
$$
\Omega^{SU(N)}_{\CN=16}
\biggr\vert_{{\bf y} \rightarrow 1;x\rightarrow 1}\;=\;\Omega^{SU(N)}_{\CN=16}\biggr\vert_{{\bf y}\rightarrow 1}\;=\;1+
\sum_{p\vert N; p\neq 1}\frac{1}{p^2}  \ ,
$$
has been computed and understood  early on \cite{Yi:1997eg,Green:1997tn,Moore:1998et}
along this line of reasoning.

The authors have also computed twisted partition functions
for more general simple groups, up to rank 4, and decomposed the resulting
$\Omega^G_{\cN}$'s in this manner\cite{Lee:2016dbm}. See
Appendix~\ref{A1} for the results. The main lesson is again that
we can read off the true Witten index $\cI$ from such
a decomposition of each $\Omega$; All the rational pieces
have to be part of $-\delta \cI$, sector by sector. The only
integral part, the first terms on the right hand sides, may
be interpreted as the Witten index, giving us
\bea
\cI^{SO(4)}_{\CN=16}=\left(\cI^{SU(2)}_{\CN=16}\right)^2 &=& 1 \ , \cr\cr
\cI^{SO(5)}_{\CN=16}=\cI^{Sp(2)}_{\CN=16} &=& 1 \ , \cr\cr
\cI^{SO(6)}_{\CN=16}=\cI^{SU(4)}_{\CN=16} &=& 1 \ , \cr\cr
\cI^{SO(7)}_{\CN=16} &=& 1 \ , \cr\cr
\cI^{Sp(3)}_{\CN=16} &=& 2 \ , \cr\cr
\cI^{SO(8)}_{\CN=16} &=& 2 \ , \cr\cr
\cI^{SO(9)}_{\CN=16} &=& 2 \ , \cr\cr
\cI^{Sp(4)}_{\CN=16} &=& 2 \ ,
\eea
as well as $\cI^{G_2}_{\CN=16} = 2 $.
In the next section, we will adopt and extend some of these
results for D-particles on an Orientifold point.

\section{D0-O0$^-$    }

Let us come to the main problem of this note.
Just as the Witten index for $\cN=16$ $SU(N)$  theory
confirms existence of M-theory circle, hidden in IIA theory,
one may ask what this M-theory circle will predict in
the presence of IIA orientifold planes. For O8 and O4,
D-particle states bound to the orientifold planes require
additional D-branes: Eight D8's for O8, since otherwise M-theory lift
does not exist \cite{Kachru:1996nd}, and more than one D4's for
O4. See Refs.~\cite{Hwang,Hwang:2016gfw} for recent computations
of twisted partition functions in the presence of O4/O8 orientifolds.
This leaves O0, namely Orientifold points. While it is, a priori,
unclear why there should be D-particles trapped at O0,
our computation of nontrivial Witten indices for $\cN=16$
$SO$ and $Sp$ theories suggests that there should be
such states after all. An orientifold projection ${\mathbb R}^9/{\mathbb Z}_2$
can give either $Sp(n)$ or $O(m)$ gauge groups. For O0$^+$'s,
the $Sp$ computation above suffices. For O0$^-$'s, however,
one must supplement $SO(m)$
computation by taking into account ${Z}_2= O(m)/SO(m)$.
In this section, we generalize $SO(m)$
to $O(m)$ theories, for D-particles bound to O0$^-$'s.

Physically, the difference between the two is whether we
demand the physical states be invariant under the gauge-parity
operation, which we call $\cP$, in addition to the local
Gauss constraint.
So if a twisted partition function for $SO(m)$ theory has the form,
$$ {\rm tr} \left[ (-1)^{\cal F}\cdots \,e^{-\beta H}\right]\ ,$$
its $O(m)$ counterpart must have the operator insertion,
$$ {\rm tr} \left[ (-1)^{\cal F}\cdots \,e^{-\beta H}\cdot\frac{1+\cP}{2}\right] \ , $$
where $\cP$ is the parity operator in $O(m)/SO(m)$.
In the end, the twisted partition function of an $O(m)$ theory
is the average of two terms,
\bea
\Omega^{O(m)}_{\cal N}({\bf y},x)=\frac{\Omega^{O^+(m)}_{\cal N}({\bf y},x) +\Omega^{O^-(m)}_{\cal N}({\bf y},x)}{2} \ .
\eea
The first term $\Omega^{O^+(m)}_{\cal N}({\bf y},x)=\Omega^{SO(m)}_{\cal N}({\bf y},x)$ has
already been computed, while the second term needs to be computed with the insertion
of $\cP$ as
\bea
\Omega^{O^-(m)}_{\cal N}({\bf y},x)\equiv  {\rm tr}
\left[ (-1)^{\cal F} {\bf y}^{R} x^{F} e^{-\beta {\cal Q}^2}\cP\right] \biggr\vert_{\rm localization}\ .
\eea

\subsection{$O(2N)$}\label{O(2N)}

First, we turn to $O(2N)$ for $2N\ge 4$.
For $\Omega^{O^-(2N)}_{\cN=4,8,16}$, we  made an
explicit JK-residue evaluation as in the previous
section. The insertion of $\cP$ can be represented
by a $Z_2$ holonomy along the Euclidean time circle,
\beq\label{P_O(2N)}
{\rm diag}_{2N\times 2N}(1,1,\dots,1,-1) \ ,
\eeq
whereby the zero mode space shrinks by one dimension,
so $r=N-1$ for $O^-(2N)$.
The reduced zero modes, $t_{1,2,\dots,N-1}=e^{2\pi i u_{1,2,\dots,N-1}}$,
parameterize $O^-(2N)$ holonomy as
$$\left(\begin{array}{ccccc}
e^{2\pi i\sigma_2 u_1}& 0& \cdots  & 0& 0\\
0& e^{2\pi i\sigma_2 u_2}& \cdots  & 0& 0\\
\cdots &  \cdots& \cdots  & 0 & 0\\
0 & 0& 0  & e^{2\pi i\sigma_2 u_{N-1}}& 0\\
0& 0& 0 & 0& \sigma_3
\end{array}\right) \ ,
$$
which sets $t_N=1$ in $g(t)$. The $N$-th Cartan
elements in all multiplets become massive, instead,
and now contribute factors with the signs flipped, e.g.,
one of the $N$ overall ${\bf y}-{\bf y}^{-1}$ factors in the denominator
for the Cartan is flipped to ${\bf y}+{\bf y}^{-1}$.
See Appendix~\ref{B}.
However, we must caution against viewing this as
a spontaneous symmetry breaking of the dynamics.
Consider very long (Euclidean) time $\beta$.
The ``symmetry breaking effect" becomes diluted arbitrarily,
as the size of the time-like gauge field scales with
$1/\beta$. Moreover, at each time slice, this $A_0$
can be gauged away, locally, and thus will not alter
the dynamics. It is only when we are instructed to
perform the trace, this $\cP$ makes a difference.

Finally, one needs to be careful about the usual
division by the Weyl group when computing $O^-(2N)$
contributions. Recall that the Weyl group of $O^+(2N)=SO(2N)$ is
\beq
W_{SO(2N)} = S_N \ltimes (\IZ_2)^{N-1} \ ,
\eeq
with the latter factor representing the even number of sign flips.
For the $O^-(2N)$ sector of the path integral, the $N$-th zero mode
is turned off and hence, the nontrivial permutation reduces to
$S_{N-1}$ while the effective number of sign-flips remains
the same. We thus need to divide by
\beq
|S_{N-1} \ltimes (\IZ_2)^{N-1}|=2^{N-1}\cdot (N-1)! \ ,
\eeq
instead of dividing by $|W_{SO(2N)}|=2^{N-1}\cdot N!$. We warn the
readers not to confuse these groups with the Weyl group of $O(2N)$
\beq
W_{O(2N)} = S_N \ltimes (\IZ_2)^{N} \ ,
\eeq
which will enter the continuum interpretation of the rational
pieces below. Just as in $O^+(2N)=SO(2N)$, the results for the
twisted partition function for $O(2N)$ can be organized physically,
in terms of plane-wave-like states that explore the classical vacua.
These plane waves will see all $N$ Cartan directions as flat,
even though in the localization computations one must regard the
$N$-th as massive. This means that the continuum contributions to
$\Omega^{O(2N)}$ will take a similar form as those to
$\Omega^{SO(2N)}$ with $W_{O(2N)}$ replacing $W_{SO(2N)}$. However,
 $W_{O(2N)}$ itself does not enter the residue computation of
 $\Omega^{O^-(2N)}_\cN$
 directly.

\subsubsection{$\cN=4,8$ }

As in section 3, we present $\cN=4,8$ results first, and
motivate how
$\cN=16$ continuum sectors should look like. This will
enable us to decompose uniquely $\cN=16$ results into the
integral part and the rational parts, in much the same
way as $\Omega^{O^+(2N)=SO(2N)}_{\cN=16}$'s were
decomposed. Having computed
$\Omega^{O^-(2N)}_{\cN=4,8}$ by a direct path integral
evaluation, we again find the results can be all organized
into the following simple expressions,
\bea\label{O-N4}
\Omega^{O^-(2N)}_{\CN=4}({\bf y})=
\frac{1}{|W_{SO(2N)}|}\sum''_{\tilde w}\frac{1}{{\rm det}\left({\bf y}^{-1}-{\bf y}\cdot \tilde w
P\right)}\ .
\eea
The sum is now over the Weyl elements of $SO(2N)$ such that
$${\rm det}\left(1-\tilde wP\right)\neq 0\ ,$$
where $P$ inside the determinant
$$P ={\rm diag}_{N\times N}(1,1,\dots,1,-1)$$
is the representation of $\cP$ on the weight
lattice of $SO(2N)$.
In this note, we will call these $\tilde w$'s
the {\it twisted} Elliptic Weyl elements.\footnote{
As an illustration, we list the first few
for $\Omega^{O^-(2N)}_{\CN=4}({\bf y})$,
\bea
\Omega^{O^-(4)}_{\CN=4}({\bf y}) &=& \frac{1}{2}\cdot\frac{1}{{\bf y}^{-2}+{\bf y}^{2}} \ , \cr\cr\cr
\Omega^{O^-(6)}_{\CN=4}({\bf y}) &=& \frac{1}{24}\left[\frac{8}{{\bf y}^{-3}+{\bf y}^3}
+\frac{1}{({\bf y}^{-1}+{\bf y})^3}\right] \ , \cr\cr\cr
\Omega^{O^-(8)}_{\CN=4}({\bf y}) &=& \frac{1}{16}\left[\frac{4}{{\bf y}^{-4}+{\bf y}^4}
 +\frac{1}{({\bf y}^{-2}+{\bf y}^2)({\bf y}^{-1}+{\bf y})^2}\right] \ .
\eea
}

Why  this happens is fairly clear in view of the heuristic
arguments in Section 3. The origin of $\Omega^{G}_{\cN=4,8}$
was understood
as a result of the orbifolding of the asymptotic Cartan
dynamics by the Weyl action, or equivalently via the
insertion of the Weyl projection operator in the Hilbert space
trace for ${\cal O}(G)$,
$$\frac{1}{|W|}\sum_{\sigma\in W} \sigma \ . $$
Only the elliptic Weyl elements $w$ with ${\rm det}(1-w)\neq 0$
contribute to $\Omega$, and produce
$$\frac{1}{|W|}\sum'_w \frac{1}{{\rm det}({\bf y}^{-1}-{\bf y}\cdot w)} \ .$$
For $O^-$'s, the operator $P$ multiplies on the
right, so the only difference is that the Weyl projection
for $O^-(2N)$ is now shifted to
$$\frac{1}{|W_{SO(2N)}|}\sum_{\sigma\in W} \sigma  P\ . $$
This leads to the modified sum (\ref{O-N4}), where $w$ is replaced by $\tilde w\cdot P$.
See Appendix~\ref{A} for more details on Elliptic Weyl elements
and {\it twisted}  Elliptic Weyl elements.

Although we computed $O^\pm(2N)$ sector contributions separately,
the total partition function
\begin{eqnarray}
\Omega^{O(2N)}_{\CN=4}({\bf y})=
\frac12\left(\Omega^{SO(2N)}_{\CN=4}({\bf y})+\Omega^{O^-(2N)}_{\CN=4}({\bf y})\right)
\end{eqnarray}
can be more succinctly written as $\Omega^{O(2N)}_{\CN=4}({\bf y})={\mathbf \Xi}^{(N)}_{\cN=4}$ with
\begin{eqnarray}\label{Xi4}
{\mathbf \Xi}^{(N)}_{\cN=4}\equiv\frac{1}{|W^{(N)}|}\sum'_w
\frac{1}{{\rm det}\left({\bf y}^{-1}-{\bf y}\cdot w \right)}\ ,
\end{eqnarray}
where the sum is now over elliptic Weyl elements of $O(2N)$
and, likewise, $W^{(N)}=W_{O(2N)}$.
This follows from the fact that $P$ is a Weyl element of
$O(2N)$ which generates $W_{O(2N)}/W_{SO(2N)}$. The universal
role played by elliptic Weyl elements is evident here again.

As in the previous section, ${\cal N}=8$ is a straightforward
extension of this, with additional factors from the single
adjoint chiral multiplet,
\bea
\Omega^{O^-(2N)}_{\CN=8}({\bf y},x)=\frac{1}{|W_{SO(2N)}|}\sum''_{\tilde w}
\frac{1}{{\rm det}\left({\bf y}^{-1}-{\bf y}\cdot \tilde w P\right)}\cdot
\frac{{\rm det}\left({\bf y}^{-1} x^{1/2}-{\bf y} x^{-1/2}\cdot \tilde w
P\right)}{{\rm det}\left( x^{1/2}- x^{-1/2}\cdot\tilde  w  P\right)}\ ,
\eea
the simplest  of which is
\bea
\Omega_{\cN=8}^{O^-(4)}({\bf y},x)=
\frac12\cdot \frac{1}{{\bf y}^{-2}+{\bf y}^2}\cdot
\frac{{\bf y}^{-2}x+{\bf y}^2x^{-1} }{x+x^{-1}}\ .
\eea
Again, we can write the total partition function as
\bea\label{Xi8}
\Omega^{O(2N)}_{\CN=8}({\bf y},x)={\mathbf \Xi}^{(N)}_{\cN=8}
\equiv\frac{1}{|W^{(N)}|}\sum'_{ w}
\frac{1}{{\rm det}\left({\bf y}^{-1}-{\bf y}\cdot  w \right)}\cdot
\frac{{\rm det}\left({\bf y}^{-1} x^{1/2}-{\bf y} x^{-1/2}\cdot  w
\right)}{{\rm det}\left( x^{1/2}- x^{-1/2}\cdot  w  \right)}\ ,
\eea
where the sum is over elliptic Weyl elements of $O(2N)$.

\subsubsection{$\cN=16$}

After computing $\Omega^{O^-(2N)}_{\cN=16}$, we again wish to decompose
it into the integral part and other rational parts from various continuum
sectors. Our findings for $\cN=4,8$ imply that there are new types of
continuum contributions that can enter $\Omega^{O^-(2N)}_{\cN=16}$, of the
form
\bea
&&\Delta^{O^-(2N)}_{\cN=16}\equiv\\\cr
&&
\frac{1}{|W_{SO(2N)}|}\sum''_{\tilde w}\frac{1}{{\rm det}\left({\bf y}^{-1}
-{\bf y}\cdot \tilde w  P\right)}\cdot
\prod_{a=1}^3
\frac{{\rm det}\left(x^{F_a/2}{\bf y}^{R_a/2-1}
-x^{-F_a/2}{\bf y}^{1-R_a/2}\cdot
\tilde wP\right)}{{\rm det}\left( x^{F_a/2}{\bf y}^{R_a/2}
- x^{-F_a/2}{\bf y}^{-R_a/2}\cdot \tilde wP\right)}\ ,\nonumber
\eea
where  the sum is over the {\it twisted} elliptic Weyl elements of $SO(2N)$.
For $\Omega^{O^-(2N)}_{\cN=16}$, we can also have continuum contributions
constructed from,
\bea
\Delta^{O^-(2r+1)}_{\cN=16}=\Delta^{SO(2r+1)}_{\cN=16} \ , \quad \text{for}~ r<N \ .
\eea
The reason for the equality is explained in next subsection.

Upon direct computations of the twisted partition functions,
the analogs of~\eqref{16} and~\eqref{otherN=16} are found
for $O^{-}(2N)$ as follows
\bea\label{OminusN=16}
\Omega^{O^{-}(4)}_{\CN=16}  &=& 1 + \Delta^{O^-(4)}_{\CN=16}\ , \nn \\ \cr
\Omega^{O^{-}(6)}_{\CN=16} &=& 1 + 3\Delta^{O^-(3)}_{\CN=16}
+ 2 \Delta^{SO(3)}_{\CN=16}\cdot \Delta^{O^-(3)}_{\CN=16}  +\Delta^{O^-(6)}_{\CN=16} \ , \nn \\ \cr
\Omega^{O^{-}(8)}_{\CN=16} &=& 2 + 2 \Delta^{O^-(3)}_{\CN=16}+ \Delta^{O^-(5)}_{\CN=16}
+ \Delta^{SO(3)}_{\CN=16}\cdot \Delta^{O^-(4)}_{\CN=16}  +\Delta^{O^-(8)}_{\CN=16} \ .
\eea
Note that the decomposition is unique.\footnote{Up to the accidental identity,
$\Delta^{O^-(2)}_{\CN=16}=2 \Delta^{O^\pm(3)}_{\CN=16}\ .$
See the subsection \ref{o2}.}
The fact that each term on the right hand side has only one
of the latter type factor is also reasonable, as at most
one subgroup $H$ would see the projection operator $\cP$.

As with $\cN=4,8$, the full partition function of
$O(2N)$ gauge theory can also be expressed in terms
of the elliptic Weyl sums,
\bea\label{Xi16}
{\mathbf\Xi}^{(N)}_{\cN=16}\equiv
\frac{1}{|W^{(N)}|}\sum'_{w}\frac{1}{{\rm det}
\left({\bf y}^{-1}-{\bf y}\cdot w\right)}\cdot
\prod_{a=1}^3
\frac{{\rm det}\left(x^{F_a/2}{\bf y}^{R_a/2-1}-x^{-F_a/2}{\bf y}^{1-R_a/2}
\cdot w\right)}{{\rm det}\left( x^{F_a/2}{\bf y}^{R_a/2}
- x^{-F_a/2}{\bf y}^{-R_a/2}\cdot w\right)}\ ,
\eea
as follows
\bea\label{ON=16}
\Omega^{O(4)}_{\CN=16}  &=& 1 + {\mathbf \Xi}^{(1)}_{\CN=16}+{\mathbf \Xi}^{(2)}_{\CN=16}\ , \nn \\ \cr
\Omega^{O(6)}_{\CN=16} &=& 1 + 2 {\mathbf \Xi}^{(1)}_{\CN=16} + {\mathbf \Xi}^{(1)}_{\CN=16}\cdot {\mathbf \Xi}^{(1)}_{\CN=16} + {\mathbf \Xi}^{(3)}_{\CN=16}\ , \nn \\ \cr
\Omega^{O(8)}_{\CN=16} &=& 2 +  3 {\mathbf \Xi}^{(1)}_{\CN=16}
+{\mathbf \Xi}^{(1)}_{\CN=16} \cdot {\mathbf \Xi}^{(1)}_{\CN=16}+2 {\mathbf \Xi}^{(2)}_{\CN=16}
+ {\mathbf \Xi}^{(2)}_{\CN=16}\cdot {\mathbf \Xi}^{(1)}_{\CN=16}  +{\mathbf \Xi}^{(4)}_{\CN=16} \ . \nn
\eea
The partition functions of $SO(2N)$ theories do
not equal those of $O(2N)$ theories,
\bea
\Omega^{O(2N)}_{\cN=16} \neq \Omega^{SO(2N)}_{\cN=16}\ ,
\eea
yet we observe that the integral pieces that enumerate threshold
bound states do agree between $O(2N)$ and $SO(2N)$,
\bea
{\cI}^{O(2N)}_{\cN=16}={\cI}^{SO(2N)}_{\cN=16} \ .
\eea
Explicit computations have shown this latter identity
for up to rank 4, and we believe this holds for all
$N$.\footnote{See section 5 for related discussions.}

\subsection{$O(2N+1)$}

One can similarly compute $\Omega^{O^-(2N+1)}_\cN$ for $N\ge 1$ via
HKY procedure, but in the end finds $\Omega^{O^-(2N+1)}_\cN = \Omega^{O^+(2N+1)}_\cN$.
Perhaps the simplest way to understand this is to
use a different form of $\cP$,
\bea
{\rm diag}_{(2N+1)\times (2N+1)}(-1,-1,\dots,-1)\ .
\eea
On representations with an even number of vector-like indices,
such as the adjoint representation or symmetric 2-tensors,
the action of $\cP$  is trivial. Neither the determinants
nor the zero modes are affected by  $\cP$, so
we find
\bea\label{OmO}
\Omega^{O(2N+1)}_{\cN} = \Omega^{SO(2N+1)}_{\cN}\ ,
\eea
for all $N$ and all $\cN=4,8,16$.
Consistent with this is the fact that the {\it twisted} elliptic Weyl
elements $\tilde w$ are in fact ordinary elliptic Weyl elements for the
case of $O(2N+1)$. This, from the trivial action of $\cP$ on
the Cartan of $SO(2N+1)$, implies that the decomposition into
continuum sectors are also intact under the projection,
leading us from (\ref{OmO}) to
\bea
{\cI}^{O(2N+1)}_{\cN}={\cI}^{SO(2N+1)}_{\cN}\ .
\eea

\subsection{$O(2)$ and $O(1)$}\label{o2}

Let us close with two exceptional cases of $O(2)$ and
$O(1)$. In the $O^+(2)=SO(2)$ sector, the
twisted partition function  vanishes
\bea\label{O2+}
\Omega^{O^+(2)}_{\cN =4,8,16}=0\ ,
\eea
as all fields are charge-neutral and the determinant
$g(t)$ is independent of the gauge variable $t$;
the relevant JK-residue sum has to vanish identically,
since we are supposed to pick up residue only from
physical poles for these pure Yang-Mills quantum mechanics \cite{Lee:2016dbm}.

For the $O^-(2)$ sector,
however, $t$ no longer appears as a zero mode, so there
is no final residue integral to perform. The localization
merely reduces to a product of determinants,
\bea
\Omega^{O^-(2)}_{\cN=4}&=&\frac{1}{{\bf y}^{-1}+{\bf y}}\;\;=\;\; 2{\mathbf \Xi}^{(1)}_{\cN=4}\, \cr\cr
\Omega^{O^-(2)}_{\cN=8}&=&\frac{1}{{\bf y}^{-1}+{\bf y}}\cdot
\frac{x^{1/2}{\bf y}^{-1}+x^{-1/2}{\bf y}}{x^{1/2}+x^{-1/2}}\;\;=\;\; 2{\mathbf \Xi}^{(1)}_{\cN=8}\, \cr\cr
\Omega^{O^-(2)}_{\cN=16}&=&\frac{1}{{\bf y}^{-1}+{\bf y}}\cdot
\prod_{a=1}^3\frac{x^{F_a/2}{\bf y}^{R_a/2-1}+x^{-F_a/2}
{\bf y}^{1-R_a/2}}{x^{F_a/2}{\bf y}^{R_a/2}+x^{-F_a/2}{\bf y}^{-R_a/2}} \;\;=\;\; 2{\mathbf \Xi}^{(1)}_{\cN=16}\,
\eea
and, in view of (\ref{O2+}),
\bea
\Omega^{O(2)}_{\cN}&=&{\mathbf \Xi}^{(1)}_{\cN}
\eea
for each $\cN=4,8,16$. Since ${\mathbf \Xi}$'s are inherently
of continuum contributions, this implies that
not only for $\cN=4,8$ but also for $\cN=16$, $\Omega^{O(2)}_{\cN}$,
the integral index vanishes,
\bea
\cI^{O(2)}_{\cN=16}=0=\cI^{SO(2)}_{\cN=16}\  .
\eea
Finally, $O(1)$ means a single D0 trapped
in O0. As such, even though the theory is empty literally,
it still makes  sense to assign,
\bea
\cI^{O(1)}_{\cN=16}=1\ ,
\eea
as the counting of a IIA quantum state. This, together with
higher rank computations above, completes $O(m)$ cases.
This result may look a little odd in that, of all orientifold
theories, the $O(2)$ theory proves to be the only case with null
Witten index. In the next section, we will explain this
from a simple and elegant M-theory reasoning.

\section{Witten Index and M-theory on ${\mathbb R}^{9}/{\mathbb Z}_2$ }

Combining results of the previous two sections, and with
help of some foresight \cite{Hanany:1999jy}, we end up with
the following, rather  compelling expressions as the generating
functions,
\bea
\label{Sp}1+\sum_{n\ge 1} z^{2n}\,\cI^{Sp(n)}_{\CN=16} &=& \prod_{k=2,4,6,\dots }(1+z^k)\ , \\\nonumber \\
\label{SO}1+\sum_{m\ge 1} z^m\,\cI^{O(m)}_{\CN=16} &=& \prod_{k=1,3,5,\dots} (1+z^k) \ .
\eea
The two generating functions count the number of
partitions of $2n$ and $m$ into, respectively, distinct
even natural numbers and distinct odd natural numbers.
Our path-integral computation confirmed this formulae
up to $2n=8$ and $m=9$, that is, up to nine D-particles
in the covering space.  Recall
that $O(2)$ is the only Orientifold theory with no bound states,
$ \cI^{O(2)}_{\cN=16}=0$. We find the manner in which (\ref{SO}) realizes
this $m=2$ result, quite compelling and elegant: $m=2$ is
the only positive integer that cannot be expressed as a
sum of distinct odd natural numbers.

A further evidence in favor of these generating functions
can be found in Ref.~\cite{Kac:1999av}, which counted
classical isolated vacua of mass-deformed theories instead.
The mass deformation is easiest to see when $\cN=16$
theory is viewed as $\cN=4$ with three adjoint chirals
and a particular trilinear superpotential $\cW$. Adding
a quadratic mass term to $\cW$, one finds certain
``distinguished" classical vacua which are cataloged
by $su(2)$ embedding, with trivial centralizers so that
the solution is isolated. Kac and Smilga proposed the
counting of such special subsets of classical vacua
equals the true Witten index of the undeformed theory.
Interestingly, this drastic approach had previously
produced the desired results of $\cI^{SU(N)}_{\cN=16}=1$
\cite{Porrati:1997ej}.

Extending this to $SO$ and $Sp$
groups, Kac and Smilga found numbers which can be seen
to be consistent with the generating functions as above.
Since $SO(m)$ theories and $O(m)$ theories are different,
one further needs to check $\cI^{SO(m)}=\cI^{O(m)}$
for all $m$, but this equality follows easily: The classical
vacua for the mass-deformed $SO(m)$ theory can be thought
of as a triplet of $m\times m $ matrices forming a $su(2)$
representation \cite{Kac:1999av}. The defining representation
of $SO(m)$ is real, so only integral spins can enter, while
the absence of centralizer demands these spins be distinct.
Each partition of $m$ into distinct odd natural numbers,
$$m=\sum k_s\ ;\qquad k_s+1\in 2{\mathbb Z}_+,\quad k_s\neq k_{s'}\;\;{\rm if}\;s\neq s'$$
then gives a solution where the three adjoints are
block-diagonal with $k_s\times k_s$ blocks.
The action of $\cP$ on such solutions is trivial, up
to possible shift along $SO(m)$ orbits, regardless of
even or odd $m$, for the same reason as $\cP$ acts
trivially on $SO(2N+1)$ pure Yang-Mills theories.

It has been observed by Hanany et. al. \cite{Hanany:1999jy} that spectrum of
type (\ref{Sp}) and (\ref{SO}) have a simple explanation in M-theory.
For this, we must first go back to the story of M-theory on
${\mathbb T}^{4p+1}/{\mathbb Z}_2$ originally due to Dasgupta
and Mukhi \cite{Dasgupta:1995zm}. $p=0$ is the well-known
Horava-Witten \cite{Horava:1995qa}, while $p=1$ is relevant for $D$-type (2,0)
theories and anomaly inflow thereof \cite{Intriligator:2000eq,Yi:2001bz,Ohmori:2014kda}. The
lesser-known case of $p=2$ was also discussed,
however, where the authors noted that the net anomaly
after the projection can be canceled by a single chiral
fermion supported at each fixed point.
As first proposed in Ref.~\cite{Hanany:1999jy}, this implies
certain spectrum of D-particle states  at the Orientifold point
${\mathbb R}^{9}/{\mathbb Z}_2$.
Upon a further ${\mathbb S}^1$ compactification, the fixed point
will become a IIA orientifold point, and at this point the
chiral fermion will generate infinite towers of harmonic
oscillators, with either integral or half-integral KK momenta,
depending on a choice of the spin structure.

With the anti-periodic spin structure, we have fermionic harmonic
oscillators $b_{k/2}, b_{k/2}^\dagger$ with odd $k$'s. The Hilbert
space built out of these, with positive KK momenta $k/2$ has
the partition function of the second type above, i.e., (\ref{SO}).
An even number of oscillators corresponds to $O(2N)$ cases of O0$^-$
while an odd number of oscillators corresponds to $O(2N+1)$ cases of
$\widetilde {\hbox{O0}^-}$.
With the periodic boundary conditions, we have
$b_{k/2}, b_{k/2}^\dagger$ with even $k$'s, instead,
so this would lead to partition function of the first
type, i.e., (\ref{Sp}). With periodic spin structure,
the zero mode $b_0, b_0^\dagger$ also appear,
meaning that there are actually two towers, built on either
the vacuum $\vert 0\rangle$ or on
$b_0^\dagger \vert 0\rangle$. It looks reasonable that we
associated these two towers with O0$^+$ and $\widetilde {\hbox{O0}^+}$,
respectively. The correspondence is complete
once we recall that $2n$ and $m$ are the D-particle
charges in the covering space and must be divided by 2.
These four towers also explain neatly the four possible types of O0's.

There are a few noteworthy facts. First, apart from the anti-D0
towers due to oscillators with negative $k$'s, there are
additional states with positive and negative $k$ oscillators
mixed. These correspond to mixture of D0 and anti-D0 from
the standard M/IIA duality, and a pair annihilation must occur
to reduce them to collection of either D0 and anti-D0 only.
The relevant coupling involves the closed string multiplet
in the bulk, as the energy must be radiated away to transverse
space. With nothing that prevents the necessary
couplings, the above four towers we reproduced from D0-O0
perspective are the only stable states from these free fermions.

Second, each of these stable states is, for any such collection
of $k$'s of the same sign, a single quantum state rather
than a supermultiplet. Although this may sound strange given
the extensive supersymmetry, there is really no contradiction
as these states are strictly one-dimensional.
Supersymmetry does not always imply an on-shell supermultiplet
for quantum mechanical degrees of freedom.
Recall that the usual D0 problem in the flat IIA case is
governed by $U(N)=U(1)\times SU(N)$, and $U(1)$ is responsible
for ${\mathbb R}^9$ center of mass degrees of freedom and the BPS multiplet
structure of 256. In the orientifold analog, this $U(1)$ is
projected out, which is consistent with the fact that O0 breaks
the spatial translational invariance completely.

Finally, the number of states at a given large D-particle quantum
number $k$ seems to grow pretty fast with $k$. For example, the
number of threshold bound states in $Sp(n)$ case equals to the number of
distinct partitions of $n$, with the known asymptotic formula \cite{asympt},
\bea
\frac{1}{4\cdot 3^{1/4}\cdot n^{3/4}}\exp\left(\pi\sqrt{n/3}\right)+\cdots \ .
\eea
This exponential growth is a straightforward consequence of
the single chiral fermion along the M-theory circle at the
origin of the IIA theory. Whether this has other physical
consequences remains to be explored.

\section{Toward Rational Invariants for Orientifolds}

For $\cN=4$  quiver theories based on $U(N)$-type gauge
groups,\footnote{This has been extensively tested in the
class of quivers where 1-cycles and of 2-cycles are absent,
meaning absence of adjoint chirals and of complex conjugate
pairs.}
it has been observed that there is a universal relationship
between $\Omega$'s and $\cI$'s of the form,
\bea\label{RI}
\Omega_\Gamma({\bf y})=
\sum_{N\vert\Gamma}
\frac{1}{N}\cdot \frac{{\bf y}^{-1}-{\bf y}}{{\bf y}^{-N}-{\bf y}^N}\cdot
\cI_{\Gamma/N}({\bf y}^N)
\eea
where the sum is over possible divisor $N$ of the
quiver $\Gamma$\cite{Lee:2016dbm}, in the sense that
$\Gamma/N$ is the same quiver except the rank vector
is divided by $N$. Not only is this structure evident in
the final answers but also in the computational middle
steps as well, and is thus quite ubiquitous in counting problems in
the wall-crossing \cite{Manschot:2010qz,Kim:2011sc,Kontsevich:2008fj}.
The object of type (\ref{RI}), prior to being identified
as the twisted partition functions \cite{Lee:2016dbm},
was also known as the rational invariants for the obvious
reason.
Note that the universal factor
\bea\label{U}
\frac{1}{N}\cdot \frac{{\bf y}^{-1}-{\bf y}}{{\bf y}^{-N}-{\bf y}^N}
\eea
in this expression coincides with $\Omega^{SU(N)}_{\cN=4}$,
and carries the continuum contribution from a plane-wave sector
of $N$-identical 1-particle-like states. This is because
the continuum sector in question resides in the Coulomb
branch, and, as such, any other $\cN=4$ $U(N)$ type quiver
theory with Coulombic flat directions can receive the
same type of contributions. Universality of this begs
for the question whether there is an analog of this
rational structure for D-brane theories with Orientifolds.

Indeed,  one of the most tantalizing
outcome is the ``orientifolded" version of (\ref{U})
$${\mathbf \Xi}_{\cN}^{(N)}$$
precisely defined in (\ref{Xi4}), (\ref{Xi8}), and (\ref{Xi16}),
as building blocks for $\Omega^G_{\cN}$ for orthogonal and
symplectic groups. These functions ${\mathbf \Xi}_{\cN}^{(N)}$
appear universally for these theories, simply because
$O(2N)$, $O(2N+1)$, and $Sp(N)$ share a common Weyl group;
$$
W_{O(2N)}=W_{O(2N+1)}=W_{Sp(N)}=W^{(N)}\equiv S_N \ltimes (\IZ_2)^{N}\ .
$$
One difference of ${\mathbf \Xi}_{\cN=4}^{(N)}$
from the above $U(N)$ version (\ref{U}) is that
${\mathbf \Xi}_{\cN}^{(N)}$ has increasing large number
of linearly independent terms, due to large number of
contributing conjugacy classes. Another complication is
that, as we saw in various $\cN=16$ Orientifolded theories,
the continuum sectors are no longer constrained to
sectors with identical partial bound states.

We note here that at least the first issue has
a simple and elegant solution; ${\mathbf \Xi}_{\cN}^{(N)}$, even though
they look individually quite complicated, can be all
constructed from a single function
${\mathbf \Xi}_{\cN}^{(1)}$.
Introducing
\bea
\chi_{\cN}^{(n)}({\bf y},\cdots )\equiv {\mathbf \Xi}_{\cN}^{(1)}({\bf y}^n, \cdots )
\eea
where the ellipsis on the left hand side denotes
other possible equivariant parameters, while the
one on the right hand side denotes the same
parameters raised to the $n$-th power,
${\mathbf \Xi}_{\cN}^{(N)}$ can be seen to be
sums of products of $\chi_{\cN}^{(n)}$ with contributing $n$'s
sum to $N$. One then finds the generating functions,
\bea
1+\sum_{N=1}^{\infty} q^N \cdot {\mathbf \Xi}_{\cN}^{(N)}
={\rm Exp}\left(\sum_{k=1}^\infty \frac{q^k}{k} \chi_{\CN}^{(k)}\right)={\rm P.E.}\left[q\cdot \chi^{(1)}_\cN\right]
\eea
for all $\cN$, where P.E. is the Plethystic Exponential \cite{Feng:2007ur}.
We expect that these quantities, term by term in $q$-expansion, should
play a role similar to (\ref{U}), now for Orientifolded quiver theories.

We are not aware of a general answer to the second complication, yet.
Trivial examples, in this sense, are $\cN=4,8$ Orientifold theories,
partition functions of which can
be paraphrased as
\bea
1+\sum_{N=1}^{\infty} q^N \Omega_{\cN =4}^{G_N}({\bf y})
={\rm P.E.}\left[\;\frac{q}{2({\bf y}^{-1}+{\bf y})}\;\right]\ ,
\eea
and
\bea
1+\sum_{N=1}^{\infty} q^N \Omega_{\cN =8}^{G_N}({\bf y},x)
={\rm P.E.}\left[\;\frac{q}{2({\bf y}^{-1}+{\bf y})}\cdot\frac{x^{1/2}{\bf y}^{-1}+x^{-1/2}{\bf y}}{x^{1/2}+x^{-1/2}}\;\right]\ ,
\eea
common for $G_N=O(2N)$, $O(2N+1)$, or $Sp(N)$. But
the analog of (\ref{RI}) for general Orientifolded
quiver theories, which may have nontrivial
ground states, is yet another matter. Even for $\cN=16$
theories computed in this note, we are yet to find a
closed form of generating functions, inclusive of all
ranks. We wish to come back to the problem of finding
generic Orientifold version of the rational invariants in near future.

\section*{Acknowledgement}

We would like to thank Chiung Hwang and Joonho Kim, for
discussions on their work involving other types of
Orientifold planes, Amihay Hanany for bringing our
attention to his old work on Orientifold points, and
Matthew Young for illuminating discussions on the quiver
stability.
SJL is grateful to Korea Institute for Advanced Study
for hospitality. The work of SJL is supported in part by
NSF grant PHY-1417316.

\appendix

\section{Elliptic Weyl Elements and Rational Invariants}\label{A}

An elliptic element $w$ of Weyl group $W$ is defined
by absence of eigenvalue 1 in the canonical
representation of $W$ on the weight lattice.

For $SU(N)$, the Weyl group $S_N$ is a little special
because the rank is actually $N-1$. The only elliptic
Weyl's are the fully cyclic ones, say, $(123\cdots N)$
and all of these belong to a single conjugacy class.
For $SO(2N)$, $SO(2N+1)$, and $Sp(N)$ groups, the Weyl
groups are $S_N$ semi-direct-product with $(Z_2)^{N-1}$,
$(Z_2)^{N}$, and $(Z_2)^{N}$, respectively. The elements
can be therefore represented as follows
$$
\sigma=(ab\dot c \dot d \dots)(klm\dot n \dots)\cdots
$$
where dots above a number indicate a sign flip.
For example $(12\dot 3)$ represents the element,
$$
\left(\begin{array}{ccc} 1 &0&0\\ 0&1&0\\ 0&0&-1\end{array}\right)\cdot
\left(\begin{array}{ccc} 0 &1&0\\ 0&0&1\\ 1&0&0\end{array}\right) \ .
$$
In this form, the above $(Z_2)^{N-1}$ for $SO(2N)$ means
that the total number of sign flip has to be even.
Since the determinant factorizes upon the above decomposition
of $w$, this should be true for each cyclic component.
It is fairly easy to see that this requires  each
cyclic component of $w$ to have an odd number of sign
flips.

Let us list the conjugacy classes of elliptic Weyl elements
for classical groups, for some low rank cases, from which
the pattern should be quite obvious,
\begin{itemize}
\item $SU(N)$
$$(123\cdots N)$$
\item $SO(4)$
$$(\dot 1)(\dot 2)$$
\item $SO(5)$ and $Sp(2)$
$$(1\dot 2),\quad(\dot 1)(\dot 2)$$
\item $SO(6)$
$$(1\dot 2)(\dot 3)$$
\item $SO(7)$ and $Sp(3)$
$$(\dot 1\dot 2\dot 3),\quad (12\dot 3),\quad (1\dot 2)(\dot 3),
\quad (\dot 1)(\dot2)(\dot 3)$$
\item $SO(8)$
$$(\dot 1\dot 2\dot 3)(\dot 4),\quad(12\dot 3)(\dot 4),\quad (1\dot 2)(3\dot 4),
\quad (\dot 1)(\dot 2)(\dot 3)(\dot 4)$$
\item $SO(9)$ and $Sp(4)$
$$(1\dot 2\dot 3\dot 4),  \quad (1 2 3\dot 4),\quad (\dot 1\dot 2\dot 3)(\dot 4),
\quad(12\dot 3)(\dot 4),\quad (1\dot 2)(3\dot 4),\quad
(1\dot 2)(\dot 3)(\dot 4), \quad (\dot 1)(\dot 2)(\dot 3)(\dot 4)$$

\end{itemize}

We may classify the twisted elliptic Weyl elements, $\tilde w$,
for $O(m)$'s, similarly. We take this to be defined by absence
of eigenvalue 1 in $\tilde w\cdot P$ where $\tilde w$ is an element
of $W_{SO(m)}$. One  immediate fact is that
the underlying action of $\cP$ is trivial on the root lattice
of $SO(2N+1)$, so for  $SO(2N+1)$, the elliptic Weyl elements
coincide with the twisted elliptic Weyl elements. This is, in retrospect,
another reason behind why $\Omega^{O^-(2N+1)}=\Omega^{O^+(2N+1)}$ and
hence $\Omega^{O(2N+1)}=\Omega^{SO(2N+1)}$.
For $O(2N)$, however, $\cP$ flips an odd number of Cartan's,

Using the same notation as above, we can then classify the
conjugacy classes of $\tilde w\cdot P$ as follows,
\begin{itemize}
\item $O^-(4)$
$$(1\dot 2)$$
\item $O^-(6)$
$$(\dot 1\dot 2\dot 3),\quad (12\dot 3),\quad (\dot 1)(\dot2)(\dot 3)$$
\item $O^-(8)$
$$(1\dot 2\dot 3\dot 4),  \quad (1 2 3\dot 4),\quad
(1\dot 2)(\dot 3)(\dot 4), $$
\item $O^-(10)$
$$(\dot 1\dot 2\dot 3\dot 4 \dot 5),  \quad (1 2 \dot 3\dot 4\dot 5),
\quad  (1 2 3 4\dot 5), $$
$$ (\dot 1\dot 2 \dot 3)(\dot 4)(\dot 5),\quad
( 1 2 \dot 3)(\dot 4)(\dot 5),
\quad (1\dot 2)(3\dot 4)(\dot 5),
\quad (\dot 1)(\dot 2)(\dot 3)(\dot 4)(\dot 5)$$
\end{itemize}
Note that $P$ is in fact nothing but the generator of
$W_{O(2N)}/W_{SO(2N)}=Z_2$. Therefore, one can also think
of $\tilde w\cdot P$ as elliptic Weyl elements of $O(2N)$
which are not in $W_{SO(2N)}$. In particular, this means that
$W_{O(2N)}=W_{O(2N+1)}=W_{Sp(N)}$ and the the respective
elliptic Weyl elements also coincide.

\subsection{$\Omega_{\cN=16}^G$  with Simple and Connected $G$}\label{A1}

We list results for twisted partition functions with $\cN=16$,
from Ref.~\cite{Lee:2016dbm};
\bea\label{otherN=16}
\Omega^{SO(4)}_{\CN=16}  &=& 1 + 2\Delta^{SO(3)}_{\CN=16}
+ \Delta^{SO(4)}_{\CN=16} \ , \\ \cr
\Omega^{SO(5)}_{\CN=16}  &=& 1 + 2\Delta^{SO(3)=Sp(1)}_{\CN=16}
+ \Delta^{SO(5)=Sp(2)}_{\CN=16}\: =\: \Omega^{Sp(2)}_{\CN=16}\ , \cr\cr
\Omega^{G_2}_{\CN=16} &=& 2 + 2\Delta^{SU(2)}_{\CN=16}
+ \Delta^{G_2}_{\CN=16} \ ,\cr\cr
\Omega^{SO(6)}_{\CN=16}  &=& 1 + \Delta^{SO(3)}_{\CN=16}
+ \Delta^{SO(6)}_{\CN=16} \ , \cr \cr
\Omega^{SO(7)}_{\CN=16} &=& 1 + 3\Delta^{SO(3)}_{\CN=16}
+  \left(\Delta^{SO(3)}_{\CN=16}\right)^2 + \Delta^{SO(5)}_{\CN=16} + \Delta^{SO(7)}_{\CN=16} \ ,\cr\cr
\Omega^{Sp(3)}_{\CN=16} &=& 2 + 3\Delta^{Sp(1)}_{\CN=16}
+ \left(\Delta^{Sp(1)}_{\CN=16}\right)^2 + \Delta^{Sp(2)}_{\CN=16} + \Delta^{Sp(3)}_{\CN=16} \ ,\cr\cr
\Omega^{SO(8)}_{\CN=16} &=& 2 + 4\Delta^{SO(3)}_{\CN=16} + 2\left(\Delta^{SO(3)}_{\CN=16}\right)^2 +
\left(\Delta^{SO(3)}_{\CN=16}\right)^3  + 3\Delta^{SO(5)}_{\CN=16} + \Delta^{SO(8)}_{\CN=16} \ ,\cr\cr
\Omega^{SO(9)}_{\CN=16} &=& 2 + 4\Delta^{SO(3)}_{\CN=16} + 2\left(\Delta^{SO(3)}_{\CN=16}\right)^2 +
2\Delta^{SO(5)}_{\CN=16}+ \Delta^{SO(3)}_{\CN=16}\cdot\Delta^{SO(5)}_{\CN=16} +\Delta^{SO(7)}_{\CN=16}
 +\Delta^{SO(9)}_{\CN=16} \ , \cr\cr
\Omega^{Sp(4)}_{\CN=16} &=& 2 + 5\Delta^{Sp(1)}_{\CN=16} + 2\left(\Delta^{Sp(1)}_{\CN=16}\right)^2 +
2\Delta^{Sp(2)}_{\CN=16}+ \Delta^{Sp(1)}_{\CN=16}\cdot\Delta^{Sp(2)}_{\CN=16} +\Delta^{Sp(3)}_{\CN=16}
 +\Delta^{Sp(4)}_{\CN=16} \ ,\nonumber
\eea
where $\Delta$'s are defined in (\ref{Delta16}).
As with $SU(N)$ case in (\ref{16}), these decompositions are unique.

\subsection{Common Building Blocks for Orthogonal and Sympletic Groups}

Since the Weyl groups of $O(2N)$, $O(2N+1)$, and $Sp(N)$
coincide,  the quantities defined in (\ref{Xi4}), (\ref{Xi8}),
and (\ref{Xi16}) are common to all three classes of the gauge
groups. These can be classified by the rank alone,
without reference to the type of orientifolding projection,
suggesting universal building blocks for continuum contributions.
Here we list a few low rank examples of ${\mathbf \Xi}^{(N)}_{\CN=4}({\bf y})$ of (\ref{Xi4});
\begin{itemize}
\item  rank 1
\bea
\frac12 \frac{1}{{\bf y}^{-1}+{\bf y}}
\eea
\item rank 2
\bea
\frac{1}{8}\left[\frac{2}{{\bf y}^{-2}+{\bf y}^2}+\frac{1}{({\bf y}^{-1}+{\bf y})^2}\right]
\eea
\item rank 3
\begin{eqnarray}
\frac{1}{48}\left[\frac{8}{{\bf y}^{-3}+{\bf y}^3}+ \frac{ 6 }{({\bf y}^{-2}+{\bf y}^2)({\bf y}^{-1}+{\bf y})}
+\frac{1}{({\bf y}^{-1}+{\bf y})^3}\right]
\end{eqnarray}
\item rank 4
\begin{eqnarray}
&&\frac{1}{384}\left[\frac{48}{{\bf y}^{-4}+{\bf y}^4}+
 \frac{32}{({\bf y}^{-3}+{\bf y}^3)({\bf y}^{-1}+{\bf y})}\right.\cr\cr
&&\hskip 1.5cm\left. + \frac{12}{({\bf y}^{-2}+{\bf y}^2)^2} +\frac{12}{({\bf y}^{-2}+{\bf y}^2)({\bf y}^{-1}+{\bf y})^2}
+\frac{1}{({\bf y}^{-1}+{\bf y})^4}\right]
\end{eqnarray}
\item rank 5
\begin{eqnarray}
&&\frac{1}{3840}\left[\frac{384}{{\bf y}^{-5}+{\bf y}^5}+
 \frac{240}{({\bf y}^{-4}+{\bf y}^4)({\bf y}^{-1}+{\bf y})}+  \frac{160}{({\bf y}^{-3}+{\bf y}^3)({\bf y}^{-2}+{\bf y}^2)} \right.\cr\cr
&&\hskip 1.5cm
+\frac{80}{({\bf y}^{-3}+{\bf y}^3) ({\bf y}^{-1}+{\bf y})^2} +\frac{60}{({\bf y}^{-2}+{\bf y}^2)^2({\bf y}^{-1}+{\bf y})}
\cr\cr
&&\hskip 1.5cm
\left.
+\frac{20}{({\bf y}^{-2}+{\bf y}^2)({\bf y}^{-1}+{\bf y})^3} +\frac{1}{({\bf y}^{-1}+{\bf y})^5}\right]
\end{eqnarray}

\end{itemize}
Elevating these to building blocks of $\cN=8,16$
orientifolded theories is a matter of attaching
chiral field contributions to each linearly-independent
rational pieces, as in (\ref{Xi8}) and in (\ref{Xi16}).
$\Omega_{\cN=4,8}$ and $\Delta_{{\cal N}=16}$'s are
related simply to these as
\bea
{\mathbf \Xi}^{(N)}_{\cN=4,8}= \Omega_{\cN=4,8}^{O(2N)}
=\Omega_{\cN=4,8}^{O(2N+1)}=\Omega_{\cN=4,8}^{SO(2N+1)}=\Omega_{\cN=4,8}^{Sp(N)} \ ,
\eea
and
\bea
{\mathbf \Xi}^{(N)}_{\cN=16}= \Delta_{\cN=16}^{O(2N)}
=\Delta_{\cN=16}^{O(2N+1)}=\Delta_{\cN=16}^{SO(2N+1)}=\Delta_{\cN=16}^{Sp(N)} \ .
\eea

\subsection{$\Omega_{\cN=16}^G$ for D-Particles on an Orientifold Point}

Although there is a universal form (\ref{Xi16}) of continuum
contributions to $\cN=16$ theories with an Orientifold point,
the actual partition functions and the indices differ among
$O(2N)$, $O(2N+1)$, and $Sp(N)$ groups. Here we list all
three series, for comparison, although $O(2N+1)$ and $Sp(N)$
cases were already shown in Section A.1 in a different notation;
\bea\label{otherN=16-O}
\Omega^{O(2)}_{\CN=16}  &=& 0+ {\mathbf \Xi}^{(1)}_{\CN=16}\ , \\ \cr
\Omega^{O(4)}_{\CN=16}  &=& 1 + {\mathbf \Xi}^{(1)}_{\CN=16}
+{\mathbf \Xi}^{(2)}_{\CN=16}\ ,  \cr \cr
\Omega^{O(6)}_{\CN=16} &=& 1 + 2 {\mathbf \Xi}^{(1)}_{\CN=16}
+ \left({\mathbf \Xi}^{(1)}_{\CN=16}\right)^2 + {\mathbf \Xi}^{(3)}_{\CN=16}\ , \cr \cr
\Omega^{O(8)}_{\CN=16} &=& 2 +  3 {\mathbf \Xi}^{(1)}_{\CN=16}
+ \left({\mathbf \Xi}^{(1)}_{\CN=16}\right)^2 +2 {\mathbf \Xi}^{(2)}_{\CN=16}
+ {\mathbf \Xi}^{(1)}_{\CN=16}\cdot {\mathbf \Xi}^{(2)}_{\CN=16}
+{\mathbf \Xi}^{(4)}_{\CN=16} \ , \cr\cr
\cr
\Omega^{O(3)}_{\CN=16}  &=& 1 + {\mathbf \Xi}^{(1)}_{\CN=16}\ , \\ \cr
\Omega^{O(5)}_{\CN=16}  &=& 1 + 2{\mathbf \Xi}^{(1)}_{\CN=16}
+ {\mathbf \Xi}^{(2)}_{\CN=16}\ , \cr\cr
\Omega^{O(7)}_{\CN=16} &=& 1 + 3{\mathbf \Xi}^{(1)}_{\CN=16}
+  \left({\mathbf \Xi}^{(1)}_{\CN=16}\right)^2
+ {\mathbf \Xi}^{(2)}_{\CN=16} + {\mathbf \Xi}^{(3)}_{\CN=16} \ ,\cr\cr
\Omega^{O(9)}_{\CN=16} &=& 2 + 4{\mathbf \Xi}^{(1)}_{\CN=16}
+ 2\left({\mathbf \Xi}^{(1)}_{\CN=16}\right)^2 +
2{\mathbf \Xi}^{(2)}_{\CN=16}+
{\mathbf \Xi}^{(1)}_{\CN=16}\cdot{\mathbf \Xi}^{(2)}_{\CN=16}
+{\mathbf \Xi}^{(3)}_{\CN=16}+{\mathbf \Xi}^{(4)}_{\CN=16} \ , \cr\cr
\cr
\Omega^{Sp(1)}_{\CN=16}  &=& 1 + {\mathbf \Xi}^{(1)}_{\CN=16}\ , \\ \cr
\Omega^{Sp(2)}_{\CN=16}  &=& 1 + 2{\mathbf \Xi}^{(1)}_{\CN=16}
+{\mathbf \Xi}^{(2)}_{\CN=16}\ , \cr\cr
\Omega^{Sp(3)}_{\CN=16} &=& 2 + 3{\mathbf \Xi}^{(1)}_{\CN=16}
+\left({\mathbf \Xi}^{(1)}_{\CN=16}\right)^2
+ {\mathbf \Xi}^{(2)}_{\CN=16} + {\mathbf \Xi}^{(3)}_{\CN=16} \ ,\cr\cr
\Omega^{Sp(4)}_{\CN=16} &=& 2 + 5{\mathbf \Xi}^{(1)}_{\CN=16}
+ 2\left({\mathbf \Xi}^{(1)}_{\CN=16}\right)^2 +2{\mathbf \Xi}^{(2)}_{\CN=16}
+ {\mathbf \Xi}^{(1)}_{\CN=16}\cdot{\mathbf \Xi}^{(2)}_{\CN=16}
+{\mathbf \Xi}^{(3)}_{\CN=16} +{\mathbf \Xi}^{(4)}_{\CN=16} \ .\nonumber
\eea

\section{Integrand for the $O^{-}(2N)$ }\label{B}

The determinant $g_{O^-(2N)}(t)$ that appears in the localization formula~\eqref{jk-formula} for the twisted partition function of the $O^-(2N)$ pure Yang-Mills theory can be obtained by modifying the following $O^{+}(2N)$ counterpart,
\bea
g_{O^+(2N)}(t)
&=&\left(\frac{1}{{\bf y}-{\bf y}^{-1}}\right)^{N}  \cdot
\prod_a \left(\frac{x^{-{F_a/ 2}}{\bf y}^{-\left({R_a/ 2}-1\right)}
-x^{{F_a/ 2}}{\bf y}^{{R_a/ 2}-1}}{
x^{{F_a/ 2}}{\bf y}^{R_a/ 2}-x^{-{F_a/ 2}}{\bf y}^{-{R_a/ 2}}}\right)^N\cr\cr
&&\times \prod_{\alpha}\frac{t^{-{\alpha/ 2}}-t^{\alpha/ 2}}{
t^{\alpha/ 2}{\bf y}^{-1}-t^{-{\alpha/ 2}}{\bf y}}  \cdot
\prod_{a}\prod_{\alpha}\frac{t^{-\alpha/2}x^{-{F_a/ 2}}{\bf y}^{-\left({R_a/ 2}-1\right)}
-t^{\alpha/2}x^{{F_a/ 2}}{\bf y}^{{R_a/ 2}-1}}{
t^{\alpha/2}x^{{F_a/ 2}}{\bf y}^{R_a/ 2}
-t^{-\alpha/2}x^{-{F_a/ 2}}{\bf y}^{-{R_a/ 2}}} \cr\cr\cr
&=&\left(\frac{1}{{\bf y}-{\bf y}^{-1}}\right)^{N} \cdot
\prod_a \left(\frac{{\bf y}^{-\left({R_a/ 2}-1\right)}
-x^{{F_a}}{\bf y}^{{R_a/ 2}-1}}{
x^{{F_a}}{\bf y}^{R_a/ 2}-{\bf y}^{-{R_a/ 2}}}\right)^N\cr\cr
&&\times \prod_{\alpha}\frac{1-t^{\alpha}}{
t^{\alpha}{\bf y}^{-1}-{\bf y}} \cdot
 \prod_{a}\prod_{\alpha}\frac{{\bf y}^{-\left({R_a/ 2}-1\right)}
-t^{\alpha}x^{{F_a}}{\bf y}^{{R_a/ 2}-1}}{
t^{\alpha}x^{{F_a}}{\bf y}^{R_a/ 2}
-{\bf y}^{-{R_a/ 2}}}  \ , \label{g_O+}
\eea
so that the parity action is appropriately taken into account.
Here, $\alpha$'s are the roots of $SO(2N)$ and $a$'s
label the $0$, $1$, and $3$ adjoint chiral multiplets for $\cN=4$, $8$, and $16$ theories, respectively.
With the parity represented as in Eq.~\eqref{P_O(2N)},
\beq
{\rm diag}_{2N\times 2N}(1,1,\dots,1,-1) \ ,
\eeq
the $N$-th zero mode is frozen to $t_N=1$ and some of the one-loop determinants relevant to the $N$-th Cartan $U(1)$ undergo appropriate sign flips as described in the paragraph including Eq.~\eqref{P_O(2N)}. The determinant $g_{O^-(2N)}(t)$ is then a function of the $N-1$ zero modes, $t=\{t_1, \dots, t_{N-1}\}$, and can be written as
\bea\label{g_O-}
g_{O^-(2N)}(t)&=&g_{O^+(2N-2)}(t) \cdot \frac{1}{{\bf y}+{\bf y}^{-1}} \cdot
\prod_a \frac{{\bf y}^{-\left({R_a/ 2}-1\right)}
+x^{{F_a}}{\bf y}^{{R_a/ 2}-1}}{
x^{{F_a}}{\bf y}^{R_a/ 2}+{\bf y}^{-{R_a/ 2}}} \\
&&\times \prod_{i=1}^{N-1}
\frac{1-t_i}{t_i{\bf y}^{-1}-{\bf y}} \; \frac{1+t_i}{t_i{\bf y}^{-1}+{\bf y}} \cdot
\prod_{i=1}^{N-1}
\frac{1-t_i^{-1}}{t_i^{-1}{\bf y}^{-1}-{\bf y}} \; \frac{1+t_i^{-1}}{t_i^{-1}{\bf y}^{-1}+{\bf y}}\;  \cr\cr
&& \times \prod_{a}\prod_{i=1}^{N-1}\frac{{\bf y}^{-\left({R_a/ 2}-1\right)}
-t_i x^{{F_a}}{\bf y}^{{R_a/ 2}-1}}{
t_i x^{{F_a}}{\bf y}^{R_a/ 2}
-{\bf y}^{-{R_a/ 2}}} \;
\frac{{\bf y}^{-\left({R_a/ 2}-1\right)}
+t_i x^{{F_a}}{\bf y}^{{R_a/ 2}-1}}{
t_i x^{{F_a}}{\bf y}^{R_a/ 2}
+{\bf y}^{-{R_a/ 2}}}\cr\cr
&& \times \prod_{a}\prod_{i=1}^{N-1}\frac{{\bf y}^{-\left({R_a/ 2}-1\right)}
-t_i^{-1} x^{{F_a}}{\bf y}^{{R_a/ 2}-1}}{
t_i^{-1} x^{{F_a}}{\bf y}^{R_a/ 2}
-{\bf y}^{-{R_a/ 2}}} \;
\frac{{\bf y}^{-\left({R_a/ 2}-1\right)}
+t_i^{-1} x^{{F_a}}{\bf y}^{{R_a/ 2}-1}}{
t_i^{-1} x^{{F_a}}{\bf y}^{R_a/ 2}
+{\bf y}^{-{R_a/ 2}}}\ , \nn
\eea
where the expression for $g_{O^+(2N-2)}(t)$ can be read from Eq.~\eqref{g_O+}.

For an illustration, we list below the determinants for the $O^-(4)$ theories with $\cN=4$, $8$, and $16$:
\bea
g_{O^-(4)}^{\cN=4}(t_1)
&=&
\frac{1}{{\bf y}-{\bf y}^{-1}} \cdot \frac{1}{{\bf y}+{\bf y}^{-1}} \cr\cr
&&\times\;
\frac{1-t_1}{t_1{\bf y}^{-1}-{\bf y}} \cdot
\frac{1+t_1}{t_1{\bf y}^{-1}+{\bf y}} \cdot
\frac{1-t_1^{-1}}{t_1^{-1}{\bf y}^{-1}-{\bf y}} \cdot
\frac{1+t_1^{-1}}{t_1^{-1}{\bf y}^{-1}+{\bf y}}\ , \cr\cr\cr
g_{O^-(4)}^{\cN=8}(t_1)
&=&
\frac{1}{{\bf y}-{\bf y}^{-1}}\cdot \frac{1}{{\bf y}+{\bf y}^{-1}}  \cdot
\frac{{\bf y}-x{\bf y}^{-1}}{x-1} \cdot \frac{{\bf y}+x{\bf y}^{-1}}{x+1} \cr\cr
&&\times\;
\frac{1-t_1}{t_1{\bf y}^{-1}-{\bf y}} \cdot
\frac{1+t_1}{t_1{\bf y}^{-1}+{\bf y}} \cdot
\frac{1-t_1^{-1}}{t_1^{-1}{\bf y}^{-1}-{\bf y}} \cdot
\frac{1+t_1^{-1}}{t_1^{-1}{\bf y}^{-1}+{\bf y}}\;\cr\cr
&&\times\;
\frac{{\bf y}-t_1 x{\bf y}^{-1}}{t_1 x-1} \cdot
\frac{{\bf y}+t_1 x{\bf y}^{-1}}{t_1 x+1} \cdot
\frac{{\bf y}-t_1^{-1} x{\bf y}^{-1}}{t_1^{-1} x-1} \cdot
\frac{{\bf y}+t_1^{-1}x{\bf y}^{-1}}{t_1^{-1}x+1}\ , \cr\cr\cr
g_{O^-(4)}^{\cN=16}(t_1)
&=&
\frac{1}{{\bf y}-{\bf y}^{-1}} \cdot \frac{1}{{\bf y}+{\bf y}^{-1}} \cdot
\frac{1-x^2}{x^2{\bf y}-{\bf y}^{-1}} \cdot \frac{1+x^2}{x^2{\bf y}+{\bf y}^{-1}}  \cr\cr
&&
\times\; \frac{{\bf y}-x^{-1}\tilde x {\bf y}^{-1}}{x^{-1}\tilde x -1} \cdot \frac{{\bf y}+x^{-1}\tilde x {\bf y}^{-1}}{x^{-1}\tilde x +1} \cdot
\frac{{\bf y}-x^{-1}\tilde x^{-1}{\bf y}^{-1}}{x^{-1}\tilde x^{-1}-1} \cdot \frac{{\bf y}+x^{-1}\tilde x^{-1}{\bf y}^{-1}}{x^{-1}\tilde x^{-1}+1} \cr\cr
&&\times\;
\frac{1-t_1}{t_1{\bf y}^{-1}-{\bf y}} \cdot
\frac{1+t_1}{t_1{\bf y}^{-1}+{\bf y}} \cdot
\frac{1-t_1^{-1}}{t_1^{-1}{\bf y}^{-1}-{\bf y}} \cdot
\frac{1+t_1^{-1}}{t_1^{-1}{\bf y}^{-1}+{\bf y}}\;\cr\cr
&&\times\;
\frac{1-t_1 x^2}{t_1 x^2{\bf y}-{\bf y}^{-1}} \cdot
\frac{1+t_1 x^2}{t_1 x^2{\bf y}+{\bf y}^{-1}} \cdot
\frac{1-t_1^{-1} x^2}{t_1^{-1}x^2{\bf y}-{\bf y}^{-1}} \cdot
\frac{1+t_1^{-1}x^2}{t_1^{-1}x^2{\bf y}+{\bf y}^{-1}}\;\cr\cr
&&\times\;
\frac{{\bf y}-t_1 x^{-1}\tilde x{\bf y}^{-1}}{t_1 x^{-1}\tilde x-1} \cdot
\frac{{\bf y}+t_1 x^{-1}\tilde x{\bf y}^{-1}}{t_1 x^{-1}\tilde x+1} \cdot
\frac{{\bf y}-t_1^{-1} x^{-1}\tilde x{\bf y}^{-1}}{t_1^{-1} x^{-1}\tilde x-1} \cdot
\frac{{\bf y}+t_1^{-1}x^{-1}\tilde x{\bf y}^{-1}}{t_1^{-1} x^{-1}\tilde x+1}\;\cr\cr
&&\times\;
\frac{{\bf y}-t_1 x^{-1}\tilde x^{-1}{\bf y}^{-1}}{t_1 x^{-1}\tilde x^{-1}-1} \cdot
\frac{{\bf y}+t_1 x^{-1}\tilde x^{-1}{\bf y}^{-1}}{t_1 x^{-1}\tilde x^{-1}+1} \cdot
\frac{{\bf y}-t_1^{-1} x^{-1}\tilde x^{-1}{\bf y}^{-1}}{t_1^{-1} x^{-1}\tilde x^{-1}-1} \cdot
\frac{{\bf y}+t_1^{-1}x^{-1}\tilde x^{-1}{\bf y}^{-1}}{t_1^{-1}x^{-1}\tilde x^{-1}+1}\ , \nn
\eea
where R-charges and flavor charges have been assigned as $R=0$ and $F=1$ to the adjoint chiral multiplet of the $\cN=8$ theory and as $R=(2,0,0)$, $F=(2,-1,-1)$ and $\tilde F = (0,1,-1)$ to the three adjoint chirals of the $\cN=16$ theory.

As a final remark, the determinant formula~\eqref{g_O-} has the following subtlety in sign. It is natural to expect that the massive Cartan factors in the first line of Eq.~\eqref{g_O-} each come with an additional minus sign,\footnote{Similar argument applies to all the flipped factors in the other lines of Eq.~\eqref{g_O-}, although the total number of such factors is always even so that they may never affect the final result.} just like they do in the $O^+$ theory,
\beq
\frac{{\bf y}^{-\left({R_a/ 2}-1\right)}
-x^{{F_a}}{\bf y}^{{R_a/ 2}-1}}{
x^{{F_a}}{\bf y}^{R_a/ 2}-{\bf y}^{-{R_a/ 2}}}
=
- \frac{x^{{F_a}/2}{\bf y}^{{R_a/ 2}-1}
-x^{-{F_a}/2}{\bf y}^{-\left({R_a/ 2}-1\right)} }{
x^{{F_a/2}}{\bf y}^{R_a/ 2}-x^{{-F_a/2}}{\bf y}^{-{R_a/ 2}}}  \ .
\eeq
If true, the formula would have an incorrect overall sign for $\cN=8$ and $16$ cases as there exist one and three such massive Cartan factors, respectively.
However, we propose that they do not come with an expected minus sign and Eq.~\eqref{g_O-} is correct as it is. For a consistency check, let us consider $\cN=4$ $O^-(2N)$ theory with an adjoint chiral multiplet, to which $R=1$ and $F=0$ are assigned.   Since this theory admits a mass term for the chiral field, it should flow to pure $\cN=4$ $O^-(2N)$ theory and hence, the twisted partition functions of the two theories must agree, with the same overall sign. We have indeed confirmed this for $N=2$ and $3$ based on the one-loop determinants~\eqref{g_O-}.

\end{document}